\documentclass[
    ,final            
  ]
  {aipproc}

\layoutstyle{6x9}

\begin{document}

\title{Extraction Of $\Delta g/g$ From Hermes Data On Inclusive Charged
Hadrons}


\author{P. Liebing, on behalf of the Hermes Collaboration}{
  address={Riken-BNL research Center, Upton, NY, 11973, USA}
}

\begin{abstract}
Hermes has measured longitudinal double spin asymmetries as a function of transverse
momentum $p_T$ using  charged inclusive hadrons from electroproduction off a
deuterium target. At  $p_T>1$ GeV, the asymmetries are sensitive to the spin
dependent gluon distribution $\Delta g$. To extract the gluon polarization
$\Delta g/g$, information on the background asymmetry and the  subprocess
kinematics has been obtained from a Leading Order Monte Carlo model. Values for 
$\Delta g/g$ have been calculated both as a function of the measured $p_T$ and
$x$, using two different methods, in the region $p_T>1.05$ GeV. 
\end{abstract}

\maketitle


\section{Introduction}
A direct, model dependent extraction of $\Delta g/g$ has been performed by Hermes
\cite{hermes1},SMC \cite{smc} and Compass \cite{compass1,compass2} for different
channels and data sets. This report presents a refined extraction  method, using
the high statistics data sample of antitagged (vetoed by electrons or positrons),
inclusive charged hadrons. A detailed study was performed to estimate the
systematic error arising within the model which uses Pythia 6.2 \cite{pythia},
and parametrizations of spin dependent parton distributions of
the nucleon and of the photon. 
\section{Experimental Data}
\begin{figure}
  \includegraphics[height=.3\textheight]
  {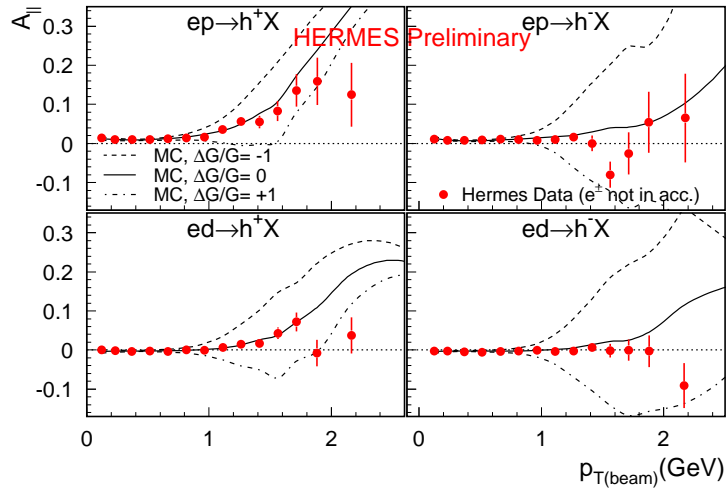}
  \caption{\label{asym_measured} Measured asymmetries for antitagged inclusive charged hadrons. The top row shows the
  asymmetries for proton- , the bottom row for deuteron target. The curves show the
  asymmetries calculated using Monte Carlo and spin dependent quark distributions and the
  assumptions $\Delta g/g(x)=-1,\,0$ and $+1$ (from top to bottom).}
\end{figure}

The data sample used for this analysis was collected using the Hermes
spectrometer. Charged inclusive hadrons were selected from events where neither a
positron nor an electron were detected. The asymmetries were calculated as
\begin{equation}
A_{\parallel}(p_T)=\frac{N^-L^+-N^+L^-}{N^-P^++N^+P^-},
\end{equation}
where $N^{+(-)}$ are the number of hadrons detected with beam and target spins
parallel (antiparallel),  $L^{+(-)}$ are the corresponding integrated
luminosities and $P^{+(-)}$ the integrated luminosities weighted with 
the product of beam and target polarizations. The
transverse momentum was calculated with respect to the beam axis. The results for
positive and negative hadrons from proton and deuteron targets are shown in
figure \ref{asym_measured}. The asymmetries have not been corrected for
acceptance and trigger efficiency. It was confirmed that the trigger efficiency does neither
introduce a significant bias to the asymmetries nor to the final results.
Also shown in the figure are the asymmetries expected from the model, using 
the assumptions  $\Delta g/g(x)=-1,\,0$ and $+1$ (lines from top to
bottom) over the full covered $x$-range. More detailed information on how the
model asymmetries were calculated will be given below. The differences
between the measured asymmetries are due to the quarks, and
are well described by the curves at low $p_T$, where the contribution from gluons
is negligible.
\section{Interpretation in Terms of the Gluon Polarization}
The measured asymmetries arise from a superposition of different subprocesses
contributing to the production of hadrons at a given measured $p_T$. In order to
decompose the asymmetries and extract the signal from processes initiated by a
hard gluon, the asymmetries and relative contributions of the background
processes have to be known as well as the hard subprocess kinematics of the
signal processes. This information was obtained from a simulation of the data
using the Pythia 6.2 Monte Carlo program and a model of the Hermes detector.  
\subsection{Data -- Monte Carlo Comparison}
The Vector Meson Dominance (VMD) Model in Pythia was adapted to reproduce the
observed exclusive $\rho^0$ cross section \cite{patty}. The fragmentation process
simulated in Jetset was tuned to multiplicities of identified hadrons at $Q^2>1$
GeV$^2$ measured at Hermes \cite{achim}. For $Q^2>0.1$ GeV$^2$, the observed
semiinclusive cross sections agree typically  within 15\% in variables {\it
integrated} over $p_T$. In contrast to this generally good agreement, the
observed cross sections do not agree vs. $p_T$. For $p_T>0.7$ GeV, the Monte Carlo
underestimates the data by factors 2 to 4. 
The disagreement is most likely due to large NLO corrections as
calculated in Ref. \cite{nlopt}.
 The LO result for $\Delta g/g$ might
therefore also be subject to large NLO corrections.
\subsection{Subprocess Fractions}
The soft background processes from
exclusive and diffractive VMD as well as nondiffractive VMD (``low-$p_T$'') are
suppressed at high $p_T$, although the ``low-$p_T$'' process still contributes
significantly for $p_T<1.5$ GeV. The quark initiated hard QCD processes together 
contribute  less than 20\% at $p_T>1$ GeV. At $p_T>1.5$ GeV, the LO DIS process
dominates the cross section. Hadrons from this process originate
predominantly from events with a large lepton scattering angle, where the
transverse momentum calculated with respect to the beam axis overestimates the transverse
momentum in the center of mass frame. The signal processes are Photon-Gluon-Fusion
(PGF) and the gluon initiated $2\rightarrow2$ (resolved photon) processes, each
contributing 10-20\% in the relevant $p_T$ range.   
\section{Extraction Methods and Results}
The background asymmetries for hard subprocesses were estimated using the MC
information on the particle types and subprocess kinematics with the nucleon PDFs
from \cite{grsv} and the photon PDFs from \cite{grs}, where the average of the
maximal and minimal scenarios was taken. The asymmetry for exclusive VMD was set
to 0, that of the ``low-$p_T$'' process was set to $A^{low-p_T}=g_1/F_1$, using
an extrapolation of a fit to world data to lower $x\approx10^{-4}$. By subtracting
the background asymmetry weighted with the background fraction from the measured
asymmetry, 
$A^{meas}_{\parallel}-R_{BG}A^{BG}=R_{sig}A^{sig}$
the signal asymmetry can be obtained. The signal asymmetry contains a convolution
of $\Delta g(x)/g(x)$ with the (polarized) hard subprocess cross section over 
the $x$-range covered by the data.  Two methods have been applied to extract the average 
$\langle\Delta g/g\rangle(p_T)$ from this asymmetry using different 
assumptions on the shape of $\Delta g(x)/g(x)$. Method I assumes that 
$\Delta g(x)/g(x)$ is essentially constant in the relevant $x$-range. Then, 
$\langle\Delta g/g\rangle(p_T)$ can be found by solving the equation
\begin{equation}
A^{meas}_{\parallel}-R_{BG}A^{BG}=R_{sig}\left\langle \hat a
\frac{\Delta f^\gamma}{f^\gamma}\right\rangle\left\langle \frac{\Delta g}{g}\right\rangle
\end{equation}
for each bin in $p_T$. Here, $\hat a$ is the hard subprocess asymmetry, and 
$\Delta f^\gamma/f^\gamma$ is the polarization of partons in the resolved
photon. 

In Method II, a functional form is assumed for $\Delta g(x)/g(x)$ which
is used to calculate the integral $A^{sig}$ for each $p_T$ bin. The functional
parameter(s) are determined by minimizing  $\chi^2$ for the difference
$A^{meas}_{\parallel}-R_{BG}A^{BG}-R_{sig}A^{sig}$
using all bins in $p_T$. 

\begin{figure}
  \includegraphics[height=.3\textheight]
  {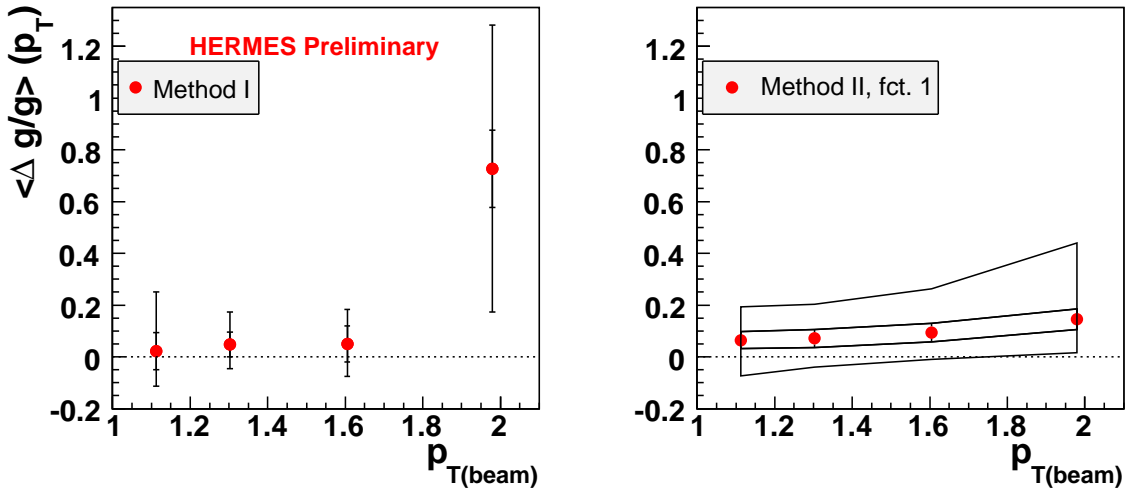}
  \caption{\label{deltag} Results for  $\langle\Delta g/g\rangle(p_T)$ from Method
  I (left) and Method II (right) for the antitagged deuterium data. The inner
  error bars (band) correspond to the statistical errors, the outer error bars
  (band) to the total errors.}
\end{figure}

Values for $\langle\Delta g/g\rangle(p_T)$ have been obtained from the deuterium
data on charge combined hadrons, for 4 bins in $p_T$ between 1.05 and 2.5 GeV.
The results are shown in figure \ref{deltag} for both Methods. Note that, in
Method II, the results and errors are correlated through the fit function.
Method I was used to confirm the overall consistency between different independent data sets
from proton and deuteron targets and positive and negative hadrons.
The experimental systematic error is approximately 14\% and arises from the
uncertainties in the beam and target polarization measurements. It is small
compared to the model uncertainty which was estimated by varying  the Pythia model
parameters, the unpolarized PDFs in the MC generation, the polarized PDFs in
the asymmetry calculation and the assumption used for the asymmetry of the 
``low-$p_T$'' process. For Method II, the functional form was also varied. 
No error was assigned on the Pythia model itself, and,
because this is a leading order approach, also no error was assigned to account
for NLO corrections.
Method II allows to determine the average $x$ of the measurement, and by
integrating over $1.05<p_T<2.5$ GeV a value of $\Delta g/g=0.071\pm 0.034 (stat)\pm 0.010
(sys-exp)^{+0.127}_{-0.105}(sys-Models)$ has been obtained 
at $\langle x\rangle = 0.22$ and $\langle\mu^2\rangle=1.35~{\rm GeV}^2$.


\bibliographystyle{aipproc}   

\bibliography{proceedings}

\IfFileExists{\jobname.bbl}{}
 {\typeout{}
  \typeout{******************************************}
  \typeout{** Please run "bibtex \jobname" to optain}
  \typeout{** the bibliography and then re-run LaTeX}
  \typeout{** twice to fix the references!}
  \typeout{******************************************}
  \typeout{}
 }

\end{document}